\documentstyle[12pt]{article}
\begin{document}
\centerline{\large\bf Four-neutrino model and the K2K experiment}
\baselineskip=8truemm
\vskip 2.5truecm
\centerline{Toshihiko Hattori$,^{a),}$\footnote{e-mail: 
hattori@ias.tokushima-u.ac.jp} \ Tsutom Hasuike$,^{b),}$\footnote{e-mail: 
hasuike@anan-nct.ac.jp} \ and \ Seiichi Wakaizumi$ ^{c),}$\footnote{e-mail: 
wakaizum@medsci.tokushima-u.ac.jp}}
\vskip 0.6truecm
\centerline{\it $ ^{a)}$Institute of Theoretical Physics, University of Tokushima,
Tokushima 770-8502, Japan}
\centerline{\it $ ^{b)}$Department of Physics, Anan College of Technology,
Anan 774-0017, Japan}
\centerline{\it $ ^{c)}$School of Medical Sciences, University of Tokushima,
Tokushima 770-8509, Japan}
\vskip 2.5truecm
\centerline{\bf Abstract}
\vskip 0.7truecm

We investigate the neutrino oscillations of $\nu_\mu$ beam at the K2K 
experiment 
in the four-neutrino model with three active and one sterile neutrinos, and 
compare them with the oscillations in the three-neutrino model. In the 
four-neutrino case, the effect of the $\Delta m^2_{\rm LSND}$ scale of  
mass-squared difference, derived from the LSND experiments, occurs in the 
survival probability $P(\nu_\mu\to\nu_\mu)$ in the range of $\Delta m^2 
< 0.004$ ${\rm eV}^2$, where $\Delta m^2$ is the relevant one to the K2K 
experiment and corresponds to the atmospheric neutrino mass scale. 
Once the probability $P(\nu_\mu\to\nu_\mu)$ is measured at the K2K, 
the allowed region of $\Delta m^2$ would turn out to be broader in 
the four-neutrino model than the one in the three-neutrino model. 
\newpage


It has turned out that neutrinos have a certain amount of mass through 
the observations of the atmospheric neutrino anomaly\cite{Kamiokande}
\cite{SuperKamiokande}. The anomaly can naturally be explained by 
the neutrino oscillation\cite{Threenu}, along with the analyses of 
the solar neutrino deficit\cite{Solar}. The oscillation, however, gives only 
the mass-squared difference among various species of neutrinos. 

At present, if the LSND experiment is included, three kinds of 
mass-squared differences are derived: $\Delta m^2_{\rm solar} = (10^{-11}
-10^{-5}){\rm eV}^2$ from the solar neutrino deficit with a large range of 
$\Delta m^2$, depending on the four solutions of the vacuum oscillation, 
and the Mikheyev-Smirnov-Wolfenstein (MSW) solutions in the matter with 
small- and large-angle mixings and the LOW one with relatively low 
mass-squared difference\cite{Bahcall}, $\Delta m^2_{\rm atm} = (1.5-5)
\times 10^{-3}$ 
${\rm eV}^2$ from the atmospheric neutrino anomaly with a large mixing 
angle of $\sin ^22\theta_{{\rm atm}}>0.82$ interpreted as the $\nu_\mu\to
\nu_\tau$ oscillation, and $\Delta m^2_{\rm LSND} = (0.2-2){\rm eV}^2$ 
from the LSND experiments on $\nu_\mu\to\nu_e$ and $\bar{\nu_\mu}\to
\bar{\nu_e}$ oscillations\cite{LSND}, which is the only one positive evidence 
from the terrestrial oscillation experiments using the accelerators and reactors. 

It is eagerly desired to know the magnitude of mass itself, which would 
be given by the neutrinoless double beta decays. It is, however, impossible 
to know the neutrino masses at present. So, for the moment, it is important 
to pinpoint 
some of the mass-squared differences to the required accuracy by doing 
various experiments. One of the experiments, which is now running, 
is the K2K experiment\cite{K2K}.

In this paper, we investigate the oscillations of the $\nu_\mu$ beam, 
which is being measured at the K2K, in the four-neutrino model with 
mass scheme of the two nearly degenerate pairs separated by the order 
of 1eV for the three active and one sterile neutrinos[8-15] by using 
the constraints 
on the mixing matrix derived from the solar neutrino deficit, atmospheric 
neutrino anomaly, Bugey reactor experiment, CHOOZ experiment, 
LSND experiments, CHORUS and NOMAD experiments and the other 
accelerator and reactor experiments. And, we compare the oscillations with 
the ones in the three-neutrino model. 

Under the neutrino oscillation hypothesis\cite{Maki}\cite{Pontecorvo}, 
the flavor eigenstates are the mixtures of mass eigenstates with mass 
$m_i$ $(i = 1, 2, 3, 4)$ in the four-neutrino model as follows: 
\begin{equation}
\nu_\alpha = \sum_{i=1}^4 U_{\alpha i}\nu_i, \qquad \alpha = e, \mu, \tau, s
\label{shiki1}
\end{equation}

\noindent
where $\nu_e, \nu_{\mu}$ and $\nu_\tau$ are the ordinary neutrinos and 
$\nu_s$ is the sterile one, and $U$ is the unitary mixing matrix. The neutrino 
oscillation probability of $\nu_{\alpha} \to \nu_{\beta}$ in vacuum is given 
in the usual manner by
\begin{equation}
P(\nu_{\alpha}\to\nu_{\beta}) = \delta_{\alpha\beta} - 4\sum_{k>j}
{\rm Re}(U^*_{\alpha k}U_{\alpha j}U^*_{\beta j}U_{\beta k})
\sin^2\Delta_{kj} + 2\sum_{k>j}{\rm Im}(U^*_{\alpha k}U_{\alpha j}
U^*_{\beta j}U_{\beta k})\sin 2\Delta_{kj},
\label{shiki2}
\end{equation}

\noindent
where $\Delta_{kj}\equiv\Delta m^2_{kj}L/(4E)$, $L$ being the distance from 
the neutrino source and $E$ the energy of neutrino. 

The four neutrino masses should be devided into two pairs of close masses 
separated by a gap of about 1eV in order to accomodate with 
the solar and atmospheric neutrino deficits and the LSND experiments 
along with the other results from the accelerator and reactor experiments 
on the neutrino oscillation. There are the following two schemes for the 
mass pattern; (i) $\Delta m^2_{{\rm solar}} \equiv \Delta m^2_{21} \ll 
\Delta m^2_{{\rm atm}} \equiv \Delta m^2_{43} \ll 
\Delta m^2_{{\rm LSND}} \equiv \Delta m^2_{32}$, and (ii) 
$\Delta m^2_{{\rm solar}} \equiv \Delta m^2_{43} \ll 
\Delta m^2_{{\rm atm}} \equiv \Delta m^2_{21} \ll 
\Delta m^2_{{\rm LSND}} \equiv \Delta m^2_{32}$, 
where $\Delta m^2_{kj} \equiv m^2_k - m^2_j$. We will 
adopt the first scheme in the following analyses, and the second scheme can 
be attained only through the exchange of indices $(1, 2)\leftrightarrow (3, 4)$ 
in the following various expressions such as the oscillation probabilities. 
The constraints on the mixing matrix $U$ are derived in the four-neutrino model 
in the following\cite{Hattori}.

(i) Solar neutrino deficit. Since $\Delta_{21} \sim 1$ and all the other five 
$\Delta_{kj}$'s are enormously larger than 1, the survival probability of 
$\nu_e$ is given from Eq.(\ref{shiki2}) by 

\begin{eqnarray}
P_{{\rm solar}}(\nu_e\to\nu_e) &\simeq& 1-4|U_{e1}|^2|U_{e2}|^2
\sin^2\Delta_{21} - 2|U_{e3}|^2(1-|U_{e3}|^2-|U_{e4}|^2)  \nonumber  \\
& & {} -2|U_{e4}|^2(1-|U_{e4}|^2) ,  \label{shiki3}
\end{eqnarray}
where the unitarity of $U$ is used. For the solar neutrino deficit, there are 
four different kinds of solutions as stated above, and a unique solution is not yet 
found, so that we will not use this deficit in order to obtain the constraints on $U$. 

(ii) Atmospheric neutrino anomaly. Since $\Delta_{21} \ll 1, \Delta_{43} 
\sim 1$ and $\Delta_{41}, \Delta_{42}, \Delta_{31}$, $\Delta_{32} 
\gg 1$, the survival probability of $\nu_{\mu}$ is given by 
\begin{equation}
P_{{\rm atm}}(\nu_{\mu}\to\nu_{\mu}) \simeq 1 - 4|U_{\mu 3}|^2
|U_{\mu 4}|^2\sin^2\Delta_{43} - 2(|U_{\mu 1}|^2+|U_{\mu 2}|^2)
(1-|U_{\mu 1}|^2-|U_{\mu 2}|^2).  \label{shiki4} 
\end{equation}
By using the data from the Super-Kamiokande experiments, 
$\sin^22\theta_{{\rm atm}}>0.82$ for $5\times 10^{-4}<
\Delta m^2<6\times 10^{-3}$ ${\rm eV}^2$, and expecting 
from this data that $|U_{\mu 1}|^2+|U_{\mu 2}|^2\ll 1$, the following 
constraint is obtained, 
\begin{equation}
|U_{\mu 3}|^2|U_{\mu 4}|^2 > 0.205 .  \label{shiki5}
\end{equation}

(iii) The Bugey experiment\cite{Bugey} (including Krasnoyarsk\cite{Krasno}, 
CDHS\cite{CDHS} and CCFR\cite{CCFR} experiments). By being typically 
represented by the Bugey reactor experiment with $L/E=3-20$ [m/MeV], 
since $\Delta_{21} \ll 1, \Delta_{43} \ll 1$ and $\Delta_{41}, \Delta_{42}, 
\Delta_{31}$, $\Delta_{32} \sim 1$, the survival probability of $\bar{\nu_e}$ 
is given by 
\begin{equation}
P_{{\rm Bugey}}(\bar{\nu_e}\to\bar{\nu_e}) \simeq 1 - 4(|U_{e3}|^2
+|U_{e4}|^2)(1-|U_{e3}|^2-|U_{e4}|^2)\sin^2\Delta_{32}.  \label{shiki6} 
\end{equation}
If we use the data from the Bugey experiment conservatively, 
$\sin^22\theta_{{\rm Bugey}}<0.1$ for $0.1<\Delta m^2<
1$ ${\rm eV}^2$, the following constraint is obtained: 
\begin{equation}
|U_{e3}|^2+|U_{e4}|^2 < 0.025 .  \label{shiki7}
\end{equation}
The first long-baseline reactor experiment, that is, the CHOOZ experiment
\cite{CHOOZ} with $L/E\sim 300$ [km/GeV] gives a constraint of 
$4|U_{e3}|^2|U_{e4}|^2 < 0.18$ through their data of $\sin^22\theta_
{{\rm CHOOZ}}<0.18$ for $3\times 10^{-3}<\Delta m^2<1.0\times 
10^{-2}$ ${\rm eV}^2$. However, this constraint can be included in the 
constraint of Eq.(7) from the Bugey experiment. 

In the same way as the above, the LSND experiments\cite{LSND} with 
$L/E=0.5-1$ [m/MeV] brings the constraint of 
\begin{equation}
|U^*_{\mu 3}U_{e3} +U^*_{\mu 4}U_{e4}| = 0.02 - 0.16    \label{shiki8}
\end{equation}
from the data of $\sin^22\theta_{{\rm LSND}}=1.5\times 10^{-3}
-1.0\times 10^{-1}$ for $0.2<\Delta m^2<2$ ${\rm eV}^2$.
And, CHORUS\cite{CHORUS} and NOMAD\cite{NOMAD} experiments
searching for the $\nu_{\mu}\to\nu_{\tau}$ oscillation with $L/E=0.02-0.03$ 
[km/GeV] give the constraint of 
\begin{equation}
|U^*_{\mu 3}U_{\tau 3} +U^*_{\mu 4}U_{\tau 4}| < 0.28    \label{shiki9}
\end{equation}
from the data of $\sin^22\theta_{{\rm NOMAD}}<0.3$ for $\Delta m^2<2.2
{\rm eV}^2$. Therefore, among the abovementioned six typical phenomena 
and experiments, the useful constraints are of Eqs. (\ref{shiki5}), (\ref{shiki7}), 
(\ref{shiki8}) and (\ref{shiki9}). 

In order to translate these four constraints to the ones for the mixing angles 
and phases, we adopt the most general parametrization of the mixing matrix 
for Majorana neutrinos, proposed by Barger, Dai, Whisnant and Young
\cite{Barger}, which includes six mixing angles and six phases. 
The expression of the matrix is too complicated to write it down here, so that 
we cite only the matrix elements which are useful for the following 
analyses; $U_{e1} = c_{01}c_{02}c_{03}$, $U_{e2} = 
c_{02}c_{03}s^*_{d01}$, $U_{e3} = c_{03}s^*_{d02}$, $U_{e4} = 
s^*_{d03}$, $U_{\mu 3} = -s^*_{d02}s_{d03}s^*_{d13} + 
c_{02}c_{13}s^*_{d12}$, $U_{\mu 4} = c_{03}s^*_{d13}$, $U_{\tau 3} = 
-c_{13}s^*_{d02}s_{d03}s^*_{d23} - c_{02}s^*_{d12}s_{d13}s^*_{d23} 
+ c_{02}c_{12}c_{23}$, and $U_{\tau 4} = c_{03}c_{13}s^*_{d23}$, 
where $c_{ij} \equiv \cos\theta_{ij}$ and $s_{dij} \equiv 
s_{ij}{\rm e}^{{\rm i}\delta_{ij}} \equiv \sin\theta_{ij}
{\rm e}^{{\rm i}\delta_{ij}}$ \cite{Barger}, and $\theta_{01}$, 
$\theta_{02}$, $\theta_{03}$, $\theta_{12}$, $\theta_{13}$, $\theta_{23}$ 
are the six angles and 
$\delta_{01}$, $\delta_{02}$, $\delta_{03}$, $\delta_{12}$, $\delta_{13}$, 
$\delta_{23}$ are the six phases. Three of the six oscillation probability 
differences are independent so that only three of the six phases determine 
the oscillation probabilities, that is, the Dirac phases. 

By using this parametrization of $U$, the four constraints of Eqs.(5), (7), 
(8) and (9) are expressed by the angles and phases as follows: 
\begin{equation}
 | -s_{02}s_{03}s_{13}{\rm e}^{-{\rm i}(\delta_{02}-\delta_{03}
+\delta_{13})}+c_{02}c_{13}s_{12}{\rm e}^{-{\rm i}\delta_{12}} |^2
c^2_{03}s^2_{13} > 0.205,       \label{shiki10}
\end{equation}
\begin{equation}
c^2_{03}s^2_{02} + s^2_{03} < 0.025,    \label{shiki11}
\end{equation}
\begin{equation}
 | c_{02}s_{02}c_{03}s_{12}c_{13} + c^2_{02}c_{03}s_{03}s_{13}
{\rm e}^{{\rm i}\delta_1}| = 0.02 - 0.16 ,  \label{shiki12}
\end{equation}
\begin{eqnarray}
&|&c^2_{02}c_{12}s_{12}c_{13}c_{23}-c_{02}s_{02}s_{03}s_{12}
c^2_{13}s_{23}{\rm e}^{-{\rm i}(\delta_1+\delta_2)}-c_{02}s_{02}
s_{03}c_{12}s_{13}c_{23}{\rm e}^{{\rm i}\delta_1}   \nonumber  \\
& & {} +c_{13}s_{13}s_{23}(c^2_{03}-c^2_{02}s^2_{12}
+s^2_{02}s^2_{03}){\rm e}^{-{\rm i}\delta_2}| < 0.28 ,   \label{shiki13}
\end{eqnarray}
where $\delta_1\equiv \delta_{02}-\delta_{03}-\delta_{12}+\delta_{13}$
and $\delta_2\equiv \delta_{12}-\delta_{13}+\delta_{23}$. 
The constraint of Eq.(\ref{shiki10}) reduces to 
\begin{equation}
s^2_{12}c^2_{13}s^2_{13} > 0.205   \label{shiki14}
\end{equation}
due to the smallness of $s_{02}$ and $s_{03}$, which is obtained from 
Eq.(11). By using the constraints of Eqs.(11) and (14) together with the nearly 
maximal mixing in the angle $\theta_{23}$, which is derived from the large 
angle mixing in $\nu_\mu\to\nu_\tau$ oscillation for the atmospheric 
neutrino anomaly, it proves that no constraints on the phases $\delta_1$ 
and $\delta_2$ are obtained from Eqs.(12) and (13). Equation (12), 
however, gives a constraint on the mixing angles. So, we obtain three 
constraints of Eqs.(11), 
(12) and (14) on the mixing angles and no constraints on the two phases of 
$\delta_1$ and $\delta_2$ in the four-neutrino model. The third Dirac phase 
does not give any significant effect to the leading parts of the oscillation 
probabilities, as can be seen from its no occurence in the constraints of 
Eqs.$(10)-(13)$. 

By using the constraints of Eqs.(11), (12) and (14), we calculate the oscillation 
probabilities of muon neutrino for the K2K experiment. We set the distance 
between the neutrino detector and the source $L$ to be 250 km and the energy 
of neutrino $E$ to be 1.4 GeV. A typical result is shown in Fig.1, which 
gives the survival probability $P(\nu_\mu\to\nu_\mu)$ and disappearance 
probabilities of $P(\nu_\mu\to\nu_\tau), P(\nu_\mu\to\nu_e)$, and 
$P(\nu_\mu\to\nu_s)$ with respect to $\Delta m^2_{43}$, which corresponds 
to $\Delta m^2_{\rm atm}$. We took the parameters as $s_{01}=s_{23}
=1/\sqrt{2}, s_{02}=s_{03}=0.11, s_{12}=0.91, s_{13}=0.67$, and 
$\delta_1=\delta_2=\pi/2$. Here and in the following, we take 
$\Delta m^2_{21} (\equiv\Delta m^2_{\rm solar}) = 1.0\times 10^{-6}$ 
${\rm eV}^2$ and $\Delta m^2_{32} ( \equiv\Delta m^2_{\rm LSND}) = 0.3$ 
${\rm eV}^2$. The difference of $P(\nu_\mu\to\nu_\mu)$ of 0.7 from 1 
around $0.0001\le\Delta m^2_{43}\le 0.001$ ${\rm eV}^2$ comes from 
the $\Delta m^2_{\rm LSND}$ contribution, as pointed out by Yasuda
\cite{Yasuda}. It goes up to $1.0-0.96$ in the same 
region of $\Delta m^2_{43}$, as the angle $s_{12}$ is increased towards 1.0. 
As seen in Fig.1, $P(\nu_\mu\to\nu_\mu)$ does not vary in $0.0001\le
\Delta m^2_{43}\le 0.001$ ${\rm eV}^2$ and decreases abruptly from 0.6 to 
0.07 in the region of $0.002\le\Delta m^2_{43}\le 0.006$ ${\rm eV}^2$. 
$P(\nu_\mu\to\nu_\tau)$ and $P(\nu_\mu\to\nu_s)$ are very small in 
$0.0001\le\Delta m^2_{43}\le 0.001$ ${\rm eV}^2$ and take a sizable 
magnitude of $0.3-0.6$ in $\Delta m^2_{43}=0.004-0.006$ ${\rm eV}^2$. 
In the computation we have not convoluted the probabilities with respect to 
the energy spread of the incident neutrinos. The feature of the computational 
results is as follows: (i) The dependence of the oscillation probabilities on 
the phases $\delta_1$ and $\delta_2$ is very weak. (ii) The change of the 
probabilities between $s_{02}=s_{03}=0.11$ and 0.05 is very small for 
$P(\nu_\mu\to\nu_\mu), P(\nu_\mu\to\nu_\tau)$ and $P(\nu_\mu
\to\nu_s)$, while the change is not so small for $P(\nu_\mu\to\nu_e)$ 
since the mixing angles $s_{02}$ and $s_{03}$ affect significantly the 
$\nu_\mu\to\nu_e$ oscillation. 
(iii) The dependence on the sign of the mixing angles $s_{12}$ and 
$s_{13}$ is that $P(\nu_\mu\to\nu_\mu)$ is same between the cases 
of $s_{12}>0, s_{13}>0$ and $s_{12}>0, s_{13}<0$, and $P(\nu_\mu
\to\nu_\tau)$ and $P(\nu_\mu\to\nu_s)$ interchange with each other 
between the two cases, and that all the probabilities do not change between 
the cases of $s_{12}>0, s_{13}>0$ and $s_{12}<0, s_{13}<0$ and also 
the same between the cases of $s_{12}>0, s_{13}<0$ and $s_{12}<0, 
s_{13}>0$. 

Next, we will  compare these results with those in the three-neutrino model 
with three active neutrinos, where the mass pattern  is taken as 
$\Delta m^2_{21}=\Delta m^2_{\rm solar}$ and $\Delta m^2_{32}
=\Delta m^2_{\rm atm}$. The constraints on the mixing matrix are derived 
as follows. The solar neutrino deficit is not used again here, since 
there are so many (four) solutions. 

(i) Atmospheric neutrino anomaly. The survival probability of $\nu_\mu$ is 
given by 
\begin{equation}
P_{{\rm atm}}(\nu_{\mu}\to\nu_{\mu}) \simeq 1 - 4|U_{\mu 3}|^2
(1-|U_{\mu 3}|^2)\sin^2\Delta_{32}.  \label{shiki15} 
\end{equation}
Then, the constraint is replaced by $|U_{\mu 3}|^2(1-|U_{\mu 3}|^2)>0.205$, 
instead of Eq.(5). This gives us the following constraint on $|U_{\mu 3}|$, 
\begin{equation}
0.54<|U_{\mu 3}|<0.84 .   \label{shiki16} 
\end{equation}

(ii) The CHOOZ experiment. The survival probability of $\bar{\nu_e}$ is 
given by 
\begin{equation}
P_{{\rm CHOOZ}}(\bar{\nu_e}\to\bar{\nu_e}) \simeq 1 - 4|U_{e3}|^2
(1-|U_{e3}|^2)\sin^2\Delta_{32}.  \label{shiki17} 
\end{equation}
The data stated in the description of the four-neutrino model gives the constraint 
of $4|U_{e3}|^2(1-|U_{e3}|^2)<0.18$, leading to the following constraint 
on $|U_{e3}|$, 
\begin{equation}
|U_{e3}|<0.22 .   \label{shiki18} 
\end{equation}
In the three-neutrino model, the constraint from the Bugey experiment can be 
included in the constraint of Eq.(18) from the CHOOZ experiment. 

(iii) The CHORUS and NOMAD experiments. The transition probability of 
$\nu_{\mu}\to\nu_{\tau}$ is given by 
\begin{equation}
P(\nu_{\mu}\to\nu_{\tau}) \simeq 4|U_{\mu 3}|^2|U_{\tau 3}|^2
\sin^2\Delta_{32}.     \label{shiki19} 
\end{equation}
Their data of $\sin^22\theta_{{\rm NOMAD}}\le 1$ for $\Delta m^2
=(1.5-5)\times 10^{-3}$ ${\rm eV}^2$ gives a constraint of 
\begin{equation}
|U_{\mu 3}||U_{\tau 3}|<0.5 .   \label{shiki20} 
\end{equation}

In order to translate these three constraints of Eqs.(16), (18) and (20) to 
the ones for the mixing angles and phases, we choose the parametrization 
used by the PDG\cite{PDG} with the additional two phases for Majorana 
neutrinos\cite{Fukuyama}, which eventually includes three angles and 
three phases. We give here only the relevant elements; $U_{e3}=s_{13}
{\rm e}^{{\rm i}(\rho -\phi)}, U_{\mu 3}=s_{23}c_{13}
{\rm e}^{{\rm i}(\rho -\beta)}$, and $U_{\tau 3}=c_{23}c_{13}$, 
where $c_{ij}\equiv \cos\theta_{ij}$ and $s_{ij}\equiv \sin\theta_{ij}$ 
and $\rho, \phi$ and $\beta$ are the three phases. As in the four-neutrino 
model, the Majorana phases $\rho$ and $\beta$ do not enter into the 
oscillation probabilities. The three constraints are expressed with the three 
angles as 
\begin{eqnarray}
0.54&<&|s_{23}c_{13}|<0.84 ,   \label{shiki21}   \\
|s_{13}|&<&0.22 ,   \label{shiki22}   \\
c^2_{13}|c_{23}s_{23}|&\le& 0.5 .   \label{shiki23}
\end{eqnarray}
The constraint of Eq.(23) is always satisfied for any values of $s_{23}$ and 
$s_{13}$. So, we have two constraints of Eqs.(21) and (22) remained. 
We again calculate the oscillation probabilities of the $\nu_\mu$ beam for 
the K2K experiment. A typical result is shown in Fig.2 for $P(\nu_\mu
\to\nu_\mu), P(\nu_\mu\to\nu_\tau)$, and $P(\nu_\mu\to\nu_e)$ 
with respect to $\Delta m^2_{32} ( = \Delta m^2_{\rm atm})$, where 
we took $s_{12}=s_{23}=1/\sqrt{2}, s_{13}=0.15$, and $\phi =\pi/2$. 
The probability $P(\nu_\mu\to\nu_\mu)$ is nearly $0.95-1$ in $0.0001\le
\Delta m^2_{32}\le 0.001$ ${\rm eV}^2$ and decreases rapidly from 0.8 
to 0.04 in the region of $0.002\le\Delta m^2_{32}\le 0.006$ ${\rm eV}^2$. 
$P(\nu_\mu\to\nu_\tau)$ takes a sizable magnitude of $0.2-0.6$ around 
$\Delta m^2_{32}=0.002-0.004$ ${\rm eV}^2$ and $P(\nu_\mu\to\nu_e)$ 
is quite small due to the small mixing angle $s_{13}$, which comes from 
the constraint by the CHOOZ experiment. The dependence of the probabilities 
on the phase $\phi$ is very weak. 

Fig.3 shows the regions of $P(\nu_\mu\to\nu_\mu)$ 
allowed by the constraints on the mixing angles and phases discussed above 
in the three-neutrino model(dashed lines) and in the four-neutrino 
model(solid lines). The allowed region is the one sandwiched between the 
upper curve(maximum) and the lower one(minimum). It  is very broad 
in the four-neutrino model, especially in the range of $0.0001\le\Delta m^2
\le 0.002$ ${\rm eV}^2$ due to the range of $0.91\le s_{12}\le 1$ derived 
from the constraint of Eq.(14), where the minimum curve of $P(\nu_\mu\to
\nu_\mu)$ is given by $s_{12}=0.91$ and the maximum one is by 
$s_{12}=1.0$. On the contrast, the region of $P(\nu_\mu\to\nu_\mu)$ 
is considerably limited in the three-neutrino model, determined by the range 
of $0.55\le s_{23}\le 0.85$ coming from the constraint of Eq.(21). 
The minimum curve is given by $s_{23}=1/\sqrt{2}$ and the maximum 
one is by $s_{23}=0.55$(or 0.85), where the maximum curve almost 
overlaps with the one in the four-neutrino model. The two models do not 
show any significant difference to $P(\nu_\mu\to\nu_\mu)$ in the range 
of $0.004\le\Delta m^2\le 0.007$ ${\rm eV}^2$, while they give a large 
difference for $\Delta m^2 < 0.004$ ${\rm eV}^2$. For example, if 
the probability $P(\nu_\mu\to\nu_\mu)$ is measured to be 0.6 by 
the K2K experiment, the three-neutrino model will predict that the relevant 
scale of mass-squared difference is $0.0030\le\Delta m^2\le 0.0034$ 
${\rm eV}^2$, a relatively narrow range, while the four-neutrino model will 
predict that it is $0.0020\le\Delta m^2\le 0.0034$ ${\rm eV}^2$, a broader 
range than in the three-neutrino model. That is to say, the determination of 
the relevant scale of $\Delta m^2$ by the K2K experiment is more flexible 
in the four-neutrino model than in the three-neutrino model. That is a big 
advantage at the time when $\Delta m^2_{\rm atm}$ is not so precisely 
determined by the experiments and/or the observations, for example, of the 
atmospheric neutrino deficit. 

A way to predict more precisely the allowed region of $P(\nu_\mu\to
\nu_\mu)$ for the K2K experiment in the four-neutrino model is to 
determine the angle $s_{12}$ to a good precision, which will be attained 
by the CHORUS/NOMAD-type experiment for the $\nu_\mu\to\nu_\tau$ 
oscillation performed with a longer distance $L \simeq 30$ km so that 
$L/E$ becomes $\simeq 1$, since $P(\nu_\mu\to\nu_\tau)$ is 
predominantly controlled by the first term with the angle $s_{12}$ 
in the left-hand side of Eq.(13).

\newpage
\centerline{\bf Figure captions}

\vskip 1.0truecm
\noindent
{\bf Fig.1.} The probabilities $P(\nu_{\mu}\to\nu_\mu)$ (solid line), 
$P(\nu_{\mu}\to\nu_\tau)$ (dash-dotted line), $P(\nu_{\mu}\to\nu_s)$ 
(dashed line), and $P(\nu_{\mu}\to\nu_e)$ (dotted line) versus 
$\Delta m^2_{43}$ calculated in the four-neutrino model for the K2K 
experiment with $L=250$ km and $E=1.4$ GeV. The parameter values of 
the mixing angles and phases are $s_{01}=s_{23}=1/\sqrt{2}, s_{02}
=s_{03}=0.11, s_{12}=0.91, s_{13}=0.67$, and $\delta_1=\delta_2
=\pi/2$. 
\vskip 0.5truecm

\noindent
{\bf Fig.2.} The probabilities $P(\nu_{\mu}\to\nu_\mu)$ (solid line), 
$P(\nu_{\mu}\to\nu_\tau)$ (dash-dotted line), and $P(\nu_{\mu}\to
\nu_e)$ (dotted line) versus $\Delta m^2_{32}$ calculated in 
the three-neutrino model for the K2K experiment. The parameter values of 
the mixing angles and phases are $s_{12}=s_{23}=1/\sqrt{2}, s_{13}
=0.15$, and $\phi =\pi/2$. 
\vskip 0.5truecm

\noindent
{\bf Fig.3.} The regions of $P(\nu_{\mu}\to\nu_{\mu})$ for the K2K 
experiment allowed by the constraints on the mixing angles and phases in 
the four-neutrino model (solid lines) and in the three-neutrino model 
(dashed lines). The allowed regions are in between the upper and lower curves. 
The upper curve in the three-neutrino model almost overlaps with the upper curve 
in the four-neutrino model. The parameter values are $s_{01}=s_{23}
=1/\sqrt{2}, s_{02}=0.024, s_{03}=0.0, s_{12}=1.0, s_{13}=0.54$ and 
$\delta_1=\delta_2=\pi/2$ for the upper curve, and $s_{01}=s_{23}
=1/\sqrt{2}, s_{02}=s_{03}=0.11, s_{12}=0.91, s_{13}=0.67$ and 
$\delta_1=0, \delta_2=\pi/2$ for the lower curve in the four-neutrino model. 
$s_{12}=1/\sqrt{2}, s_{23}=0.55, s_{13}=0.20$ and $\phi =\pi/2$ 
for the upper curve, and $s_{12}=s_{23}=1/\sqrt{2}, s_{13}=0.20$ 
and $\phi =\pi/2$ for the lower curve in  the three-neutrino model. 
\vskip 0.5truecm

\end{document}